\documentstyle[aps,prl,multicol,epsf]{revtex} 
\newcommand{\fig}[2]{
  \begin{figure}[t]
      \centerline{\epsfxsize=3.1in \epsffile{#1}}   
      \vspace{0.5ex}
      \caption[]{\label{#1} #2}
  \end{figure}
}

\begin{document}
\draft
\title{Supershells in Metal Clusters:\\
       Self-Consistent Calculations and their Semiclassical Interpretation}
\author{Erik Koch}
\address{Max-Planck-Institut f\"ur Festk\"orperforschung, D-70569 Stuttgart}
\date{28 September 1995}
\maketitle 

\begin{abstract}
To understand the electronic shell- and supershell-structure in large
metal clusters we have performed self-consistent calculations in the
homogeneous, spherical jellium model for a variety of different
materials. A scaling analysis of the results reveals a surprisingly
simple dependence of the supershells on the jellium density.
It is shown how this can be understood in the framework of a
periodic-orbit-expansion by analytically extending the well-known
semiclassical treatment of a spherical cavity to more realistic
potentials.  
\end{abstract}
\pacs{36.40.Cg, 31.15.Ew, 31.15.Gy\hfill cond-mat/9606023} 

\vspace{-2ex}
\begin{multicols}{2}
\narrowtext
The structure observed in the mass spectra of large, warmed metal
clusters \cite{cpl91,nature91,Brechignac93,Broyer93} can be attributed
to the properties of itinerant electrons moving in a finite volume
\cite{nishioka90,GenzkenPRL}. The most prominent finite-size effect is
the occurrence of pronounced oscillations in the density of states
\cite{BaBlo3} giving rise to an oscillating part $\tilde{E}$ of the
total energy, which is superimposed on the smooth Thomas-Fermi energy
$\bar{E}$. With increasing cluster radius, one finds regular oscillations 
({\em shells}) whose amplitude is modulated ({\em supershells}).

Two different theoretical approaches have been used to describe the
electronic shell- and supershell-structure in large metal clusters. One
is the self-consistent jellium model \cite{ekardt84}. In its simplest
form a cluster is described by a homogeneous sphere of given charge
density, dressed with $N$ valence electrons. Treating the electrons
self-consistently using density functional theory, the only input
parameter for such a calculation is the Wigner-Seitz radius $r_s$.
Although describing the electronic structure of alkali clusters quite well, 
this model provides little physical insight into the mechanisms underlying the 
shell- and supershell oscillations. It is here that the second
approach comes in. Given an effective one-particle potential one can
find a semiclassical expansion of the oscillating part of the density
of states in terms of classical periodic orbits
\cite{BaBlo3,Gutzwiller70}. Introducing a suitable damping factor one
finds for the spherical cavity, that the oscillations in the density of
states are essentially determined by the contributions of triangular
and square orbits. The supershells can thus be understood as a beating
pattern originating from the contributions of these orbits.

The semiclassical approach draws its power from the fact that the
periodic-orbit-expansion is known analytically for the model case of
the spherical cavity. However, potentials with hard walls are
only a crude approximation to  realistic cluster potentials, which have a
soft surface. Such potentials can also be treated 
using semiclassical techniques \cite{Lerme93a,Lerme93b}. In particular 
ultra-soft potentials have received much attention \cite{Manninen93,Lerme92}. 
Unfortunately in these cases 
the action integrals entering the semiclassical
formalism have to be evaluated numerically.

In order to find out what determines the shell- and supershell-structure in
metal clusters, we have carried out the following program. First we performed
a series of calculations in the homogeneous, spherical jellium model 
for a range of different electron densities. A scaling analysis 
of the results suggested that changing the electron density merely introduces 
a phase shift in the supershell structure. It is shown how this can be 
understood semiclassically in terms of a {\em leptodermous expansion}, where 
the action integrals for the potential under consideration are expanded 
around a cavity. Thereby we obtain an {\em analytical} expression for
the shift of the supershells. Finally it is shown that the leptodermous 
expansion works for realistic cluster potentials by comparing the shifts 
of supershells extracted from self-consistent calculations to those given 
by the semiclassical formulae.

As starting point of our analysis, we have performed extensive
calculations in the homogeneous, spherical jellium model. 
We use the local density approximation in the parameterization given in
Ref.\ \cite{VWN_LDA}.
Electron densities range from $r_s=2.07\;a_0$ for aluminum to $r_s=5.63\;a_0$,
corresponding to bulk cesium. Cluster sizes were chosen from $N=100$
to $6000$ valence-electrons, thus including the first two nodes of the
supershell oscillation for all densities considered. 
Typical results for $\tilde{E}(N)$ are shown in fig.\ \ref{jellyosc.eps}.  
It can be seen that the supershells are shifted towards larger $N$ as the 
Wigner-Seitz radius decreases, while the positions of the shell-minima are 
fairly independent of $r_s$. To quantify the supershells we determine the 
envelope of $\tilde{E}(N)$ by low-pass filtering its absolute value.
The position of the supershell-nodes are given by
the minima in the envelope. Filtering out the shell structure of
course introduces an uncertainty of the order of the distance between
adjacent shell-minima. The results of our jellium calculations are 
listed in table~\ref{tab_snodepos}.

To make a quantitative comparison of the different jellium
calculations, we describe the problem in terms of dimensionless
quantities. In order to do so, the relevant scales of the problem have to
be identified. Obviously one such scale is the 
Wigner-Seitz radius. In fact, we find that the amplitude of the oscillations 
$\tilde{E}(N)$ are 
 \fig{}{
   Oscillating part of the total energy extracted from self-consistent
   calculations in the homogeneous, spherical jellium model. The positions
   of the super-nodes are indicated.}

\noindent
proportional to $1/r_s^2$. The existence of a surface 
introduces an additional scale: the width $a$ of the surface region. 
As has been shown in \cite{smith69}, $a$ is fairly independent of $r_s$.  
Assuming the shift of the supershells to be a
surface-effect, we identify $a/r_s \propto 1/r_s$ as the relevant
scaling parameter. By plotting the positions of the super-nodes as a
function of $1/r_s$ (cf.~fig.~\ref{jellynodl.eps}) we indeed find a simple 
relation: The super-nodes are linearly shifted as a function 
of $1/r_s$. In particular, the first and second super-node are shifted 
{\em in parallel}. Describing $\tilde{E}(N)$ semiclassically as a simple 
beating pattern \cite{nishioka90}, it is therefore tempting to conclude 
that the shift of the supershells is caused by a phase-shift in the 
contributions of the periodic orbits.

 \noindent
 \begin{table}
 \begin{tabular}{lddd}
   material & $r_s$ in $a_0$ & 1st super-node & 2nd super-node \\ \hline
   Cs & 5.63 &  8.39 & 14.59 \\
   Rb & 5.20 &  8.43 & 14.67 \\
   K  & 4.86 &  8.47 & 14.75 \\
   Na & 3.93 &  8.95 & 15.11 \\
   Li & 3.26 &  9.15 & 15.45 \\[1ex]
   Tl & 2.48 &  9.79 & 16.01 \\
   In & 2.41 &  9.85 & 16.11 \\
   Ga & 2.19 & 10.13 & 16.37 \\
   Al & 2.07 & 10.20 & 16.49
 \end{tabular}
 \caption[]{\label{tab_snodepos}
   Position of super-nodes (given as $N^{1/3}$) for different 
   jellium-densities.}
 \end{table}

To check this conjecture, we derive an explicit periodic-orbit-expansion 
for the oscillating part $\tilde{E}(N)$ of the total energy. 
 Our approach is based on the fact that $\tilde{E}$ can, to first order,
 be extracted from the spectrum of smooth potentials that fit the 
 self-consistent results \cite{HarrisArgument}.
We start from the observation \cite{BerryMount} that in the semiclassical 
approximation the density of states naturally separates
into two contributions: the smooth Thomas-Fermi term $\bar{\rho}$ and
an oscillating contribution $\tilde{\rho}$. The latter term describes 
the quantum corrections to Thomas-Fermi theory and is given by an 
expansion over all classical periodic orbits which exist in the potential 
under consideration. For a spherically symmetric potential with exactly two
classical turning points the periodic orbits can be uniquely labeled by two 
positive integers: the number of times $\lambda$ it turns around the origin,
and the number $\nu$ of vertices it has. Denoting the classical action along
such a periodic orbit by $S_{(\lambda,\nu)}$, one finds an expression
of the form \cite{BaBlo3}:
\begin{equation}\label{DOSPOE}
\tilde{\rho}(E)\,dE = \sum_{(\lambda,\nu)} A_{(\lambda,\nu)}
    \cos\Big(S_{(\lambda,\nu)}/\hbar - \varphi_{(\lambda,\nu)}\Big)\;dE,
\end{equation}
where $\varphi_{(\lambda,\nu)}$ is the so called Maslov-phase.
Unfortunately the expansion (\ref{DOSPOE}) 
converges quite slowly. This is obvious, since it is supposed to 
approximate the density of states, which is a sum of $\delta$-functions. 
Therefore 
one usually introduces some damping as to broaden the eigenstates and make 
the expansion (\ref{DOSPOE}) converge more rapidly. 
But we are actually not interested in the density of states, rather
in $\tilde{E}(N)$. In the limit of large $N$ (which corresponds to the
semiclassical limit) the oscillating part of the total energy is given by
\begin{equation}
\tilde{E}(N) = -\int_0^{\bar{E}_F(N)}dE 
                  \int_0^E dE'\,\tilde{\rho}(N;E').
\end{equation}
 \fig{}{
   Position of the first and second super-node as a function of $1/r_s$.
   The error bars are due to the uncertainty in locating the super-nodes
   as minima of the envelope of $\tilde{E}(N)$. The solid line gives a
   linear fit to the data, the parameters of which are given in the plot.}
Here $\bar{E}_F(N)$ is the Fermi energy in Thomas Fermi approximation.
Integrating twice by parts over $\tilde{\rho}$ essentially divides the 
expansion parameters  $A_{(\lambda,\nu)}$ in (\ref{DOSPOE}) by the square 
of the classical action along the orbit, thereby reducing the importance 
of the longer orbits: 
\begin{equation}\label{EoscPOE}
\tilde{E}(N)= \bar{E}_F^2
  \sum_{(\lambda,\nu)} {4 A_{(\lambda,\nu)}\over S_{(\lambda,\nu)}^2}
    \cos\Big(S_{(\lambda,\nu)}/\hbar - \varphi_{(\lambda,\nu)}\Big).
\end{equation}
Thus there is no need of introducing a damping factor to accelerate 
convergence.  Actually, 
$\tilde{E}(N)$ is dominated by the contributions of the shortest plane
periodic orbits, namely the triangular and the square orbit. Furthermore
inspection of (\ref{EoscPOE}) shows that variations in the boundary conditions, 
which strongly shift the oscillations in $\tilde{\rho}$ \cite{Tatievski94}, 
hardly influence $\tilde{E}$, since the changes in the density of states are 
compensated by those in the Fermi energy.

As an immediate application of the expansion (\ref{EoscPOE}), we can
investigate how an increase in density for small clusters compared to 
the bulk affects the electronic shells and supershells. Such a 
{\em lattice contraction} was suggested by EXAFS analyses of small clusters
\cite{EXAFS1}. Changing the density clearly will change the Fermi-energy
for a given cluster. As we can see from equation (\ref{EoscPOE}) this will 
obviously change the overall amplitude of $\tilde{E}(N)$. But apart from
that the oscillations are determined by the classical actions 
$S_{(\lambda,\nu)}$ along the periodic orbits. For a spherical cavity of
radius $R_0$ we find
\begin{equation}\label{caviact}
  S_{(\lambda,\nu)}/\hbar = 2\nu\,\sin(\pi\lambda/\nu)\; \bar{k}_F R_0
  \;;\quad \bar{k}_F=\sqrt{2m\bar{E}_F}/\hbar ,
\end{equation} 
where the product $\bar{k}_F R_0$ depends on the number $N$ of electrons
inside the cavity, but is independent of the electron-density. Hence the
electronic shell-structure for spherical-cavity-clusters does not depend
on any lattice contraction, except for an overall change in amplitude. 
This result suggests that the same is true for smooth potentials, provided
the lattice contraction is not too large. We have confirmed this by 
numerically solving the quantum mechanical problem for realistic potentials
introducing contractions $\Delta R_0$ of up to $1/2\;r_s$.
Thus we can conclude that a possible lattice contraction will not noticeably
affect the electronic shells and supershells.

Next we turn to the problem of understanding why the super-nodes are 
phase-shifted as a function of jellium density (cf.~fig.\ 
\ref{jellynodl.eps}). The most straightforward approach would be to solve
the integrals, which enter equation (\ref{EoscPOE}), explicitly. 
Unfortunately, this cannot be done analytically. But in the semiclassical 
limit, which corresponds to $N\!\to\!\infty$, it is sufficient to know the
integrals to leading order in $1/N$. The basic idea is then to use the 
spherical cavity as a starting point and expand the action for more 
realistic potentials around this case, using the surface-width $a$ as 
small parameter ({\em leptodermous expansion}). For the classical 
action we then find 
\begin{equation}\label{Sexpand}
S_{(\lambda,\nu)}/\hbar = S^{cavity}_{(\lambda,\nu)}/\hbar
  + \Big(I_1 + I_2{a\over r_s}\Big) + {\cal O}(N^{-1/3}),
\end{equation} 
where the expansion parameters $I_1$ and $I_2$ are independent of
cluster-size; i.e.\ to leading order surface softness introduces a 
phase-shift in the periodic-orbit-expansion (\ref{EoscPOE}),
while the period of the oscillations is still determined by
$S^{cavity}_{(\lambda,\nu)}/\hbar$. An additional phase shift 
arises from the difference of the Maslov phases $\varphi_{(\lambda,\nu)}$
for a soft potential or a cavity. Finally an inspection
of the amplitudes in (\ref{EoscPOE}) shows that, to leading
order in $1/N$, they do not dependent on the shape of the potential.
We thus find that replacing a cavity by a soft potential
with small surface-parameter $a$ amounts to merely shifting phases
in the periodic-orbit-expansion of $\tilde{E}(N)$.

Now the question arises, whether typical cluster potentials are
such that their surface-parameter is small enough for the above expansion
to be valid. To judge this, we have to fit the potentials obtained from
our self-consistent calculation with some analytical model potential.
Since the classical action depends only on the potential in the classically
allowed region, it seems reasonable to fit only for $E<E_F$. There the 
self-consistent potential, except for possible Friedel oscillations, can be 
well described by a Woods-Saxon function
\begin{equation}\label{WSpot}
V(r) = {-V_0 \over 1 + \exp\left({r-R_0\over a}\right)} .
\end{equation}
But fitting only for $E<E_F$ seems to imply an error in
calculating the Maslov phases which serve to capture the influence of the 
classically forbidden region. From WKB quantization it can be seen  
that the Maslov phases for a separable system are given by the sum of the 
quantum-mechanical scattering-phases at the classical turning points. For 
Woods-Saxon potentials we can calculate these analytically. To leading order, 
they coincide with the Maslov phases for a square-well potential of depth 
$-V_0$ \cite{Tatievski94}. Thus, the error in the potential 
for $E>E_F$ will not enter the leptodermous expansion. 

Given the potential (\ref{WSpot}), we have found analytical expressions 
for the parameters $I_1$ and $I_2$ in the expansion (\ref{Sexpand}) of the 
classical action. Introducing the abbreviations
\begin{displaymath}
P   = \sqrt{E_F+V_0\over V_0}  \quad\mbox{ and }\quad
P_L = P\cos\left({\pi\lambda\over\nu}\right),
\end{displaymath}
the expansion parameters for a given periodic orbit $(\lambda,\nu)$ are 
\begin{displaymath}
  I_1 = {3\nu\over2}\sin\left({\pi\lambda\over\nu}\right)
      \left[\left({1\over P^2}-2\right)\arcsin(P)-\sqrt{{1\over P^2}-1}\;\right]
\end{displaymath} 
and
\begin{eqnarray}
 && I_2= 
 4\nu\left({9\pi\over4}\right)^{1\over3} 
    \Bigg[{P_L\over P}\ln(2P_L)-\sqrt{1-P_L^2 \over P^2}\arcsin(P_L) \nonumber\\
     & &    -\sin\left({\pi\lambda\over\nu}\right)\left(\ln(2P)
                       +\left({1\over P^2}
                       -1\right)^{3\over2}\arcsin(P)-{1\over P^2}\right)
       \Bigg]. \nonumber 
\end{eqnarray}
Details of the calculation will be published elsewhere \cite{ToBePublished}.

We are now in the position to find a simple estimate for the shift of the 
supershells. Following \cite{nishioka90} we start from a drastically 
simplified version of equation (\ref{EoscPOE}): All periodic orbits except 
triangular $(1,3)$ and the square $(1,4)$ orbits are neglected. Furthermore
it is assumed that the amplitudes for these orbits are equal.
This leaves us with an expression of the form
\begin{equation}\label{beating}
  \tilde{E} = A \left(  \cos\big(f_1\,N^{1\over3} + \varphi_1\big)
                      + \cos\big(f_2\,N^{1\over3} + \varphi_2\big)\right).
\end{equation} 
Fitting the self-consistent potentials with (\ref{WSpot}), we can
compare the shift observed in the jellium calculations to that determined
by the leptodermous expansion using the ansatz (\ref{beating}). 
This is shown in fig.~\ref{Ntcomparel.eps}.%
\fig{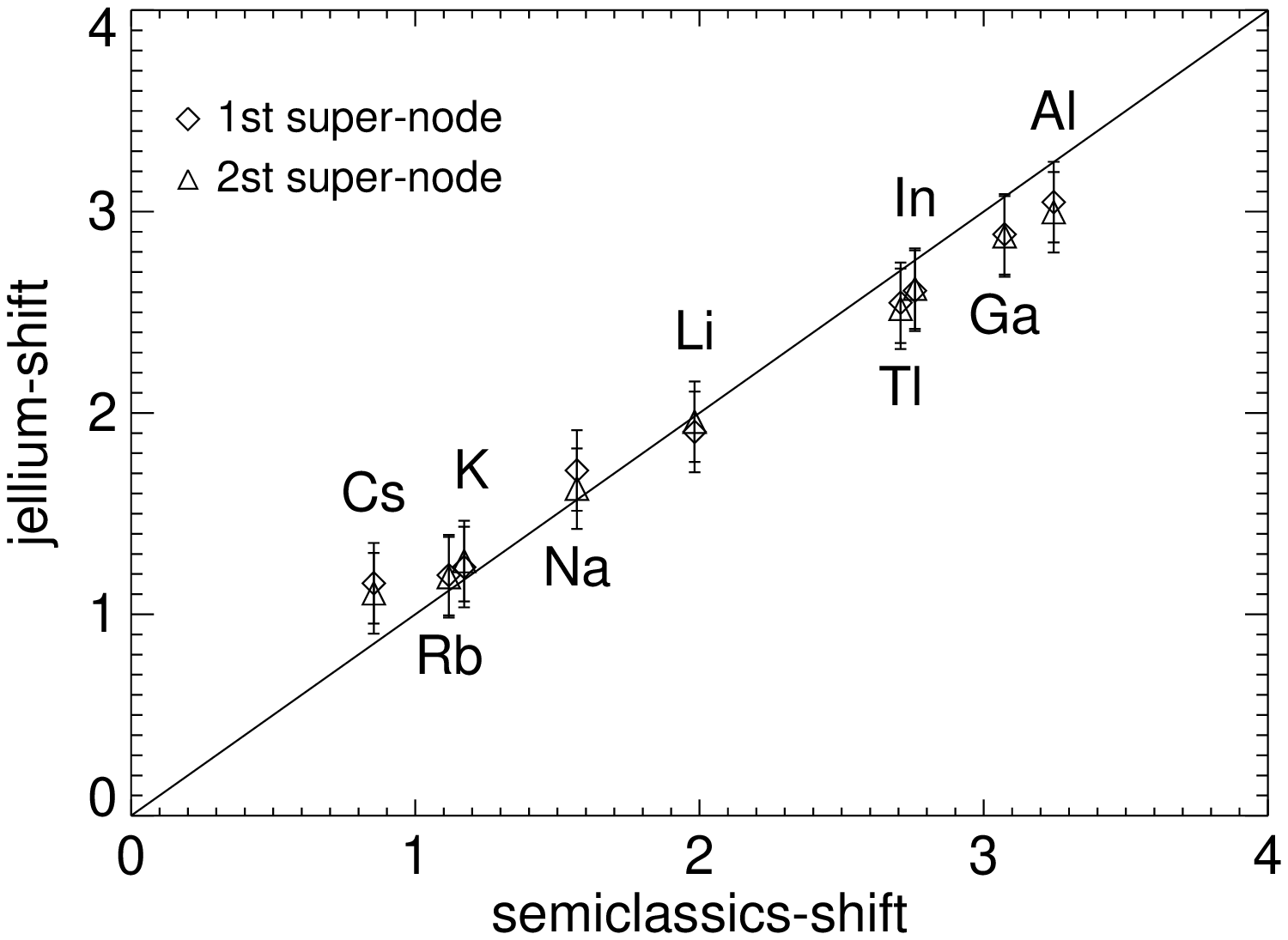}{
  Shift of super-nodes (in units of $N^{1/3}$) for different jellium clusters.
  The shift obtained from the periodic orbit expansion is compared to that
  observed in the self-consistent calculations. If these shifts agreed 
  perfectly, the points would fall on the full line.}
Although we have introduced a number of approximations the agreement is 
remarkable. 
We can thus conclude that the leptodermous expansion applies to typical 
cluster potentials.

To summarize, we have demonstrated how the semiclassical description of 
a spherical cavity can be generalized to describe the electronic supershells
of realistic potentials. Starting from a periodic-orbit-expansion for 
the oscillating part of the total energy, we have shown that $\tilde{E}(N)$ 
is hardly influenced by the Maslov Phases or a lattice contraction. Introducing
a leptodermous expansion for the classical action we have established that 
a soft potential gives phase-shifts in the semiclassical expression
for $\tilde{E}(N)$. We can thus understand the
dependence of the supershell on the electron density, revealed by a scaling 
analysis of our jellium calculations.

Moreover the leptodermous expansion can be used to analyze how the 
supershell structure will change if the underlying model is changed.
Introducing a pseudopotential, as e.g.\ in the stabilized jellium model
\cite{StabilJelly}, increases the spill-out of the electrons and leads
to a softening of the potential at the surface of the cluster. This induces 
a shift of the super-nodes towards larger $N$, which can be estimated by
the semiclassical technique described above. 


Finally, the identification of $r_s$ as the typical length scale for
the supershell-problem suggests a justification of the ad-hoc procedure
proposed in \cite{Broyer93} to improve the results of jellium calculations 
for gallium clusters. There it was found that the introduction of a
non-homogeneous jellium background is essential for treating $Ga_N$
clusters, while alkali clusters are well described by a homogeneous
jellium. Assuming, that the typical length scale for features in the
jellium is the ionic radius $r_{at}$, while the length scale for the electrons
is the Wigner-Seitz radius $r_s$, we find that the importance of 
inhomogeneities increases with the number of
valence electrons $Z_{val} \propto (r_{at}/r_s)^3$.

I am much indebted to O.\ Gunnarsson for his invaluable advice.
Helpful discussions with T.~P.~Martin~and M.~Brack are gratefully acknowledged.

\bibliographystyle{prsty}

\begin{thebibliography}{10}

\bibitem{cpl91}
T.~P. Martin {\it et~al.}, Chem.\ Phys.\ Lett.\ {\bf 186},  53  (1991).

\bibitem{nature91}
J. Pedersen {\it et~al.}, Nature (London) {\bf 353},  733  (1991).

\bibitem{Brechignac93}
C. Br\'echignac {\it et~al.}, Phys.\ Rev.\ B {\bf 47},  2271  (1993).

\bibitem{Broyer93}
M. Pellarin {\it et~al.}, Phys.\ Rev.\ B {\bf 48},  17645  (1993).

\bibitem{nishioka90}
H. Nishioka, K. Hansen, and B.~R. Mottelson, Phys.\ Rev.\ B {\bf 42},  9377
  (1990).

\bibitem{GenzkenPRL}
O. Genzken and M. Brack, Phys.\ Rev.\ Lett.\ {\bf 67},  3286  (1991).

\bibitem{BaBlo3}
R. Balian and C. Bloch, Ann.\ Phys.\ (N.Y.) {\bf 69},  76  (1972).

\bibitem{ekardt84}
W. Ekardt, Phys.\ Rev.\ B {\bf 29},  1558  (1984).

\bibitem{Gutzwiller70}
M.~C. Gutzwiller, J.\ Math.\ Phys.\ {\bf 11},  1791  (1970).

\bibitem{Lerme93a}
J. Lerm\'e {\it et~al.}, Phys.\ Rev.\ B {\bf 48},  9028  (1993).

\bibitem{Lerme93b}
J. Lerm\'e {\it et~al.}, Phys.\ Rev.\ B {\bf 48},  12100  (1993).

\bibitem{Manninen93}
J. Mansikka-aho and M. Manninen, Phys.\ Rev.\ B {\bf 48},  1837  (1993).

\bibitem{Lerme92}
J. Lerm\'e {\it et~al.}, Phys.\ Rev.\ Lett.\ {\bf 68},  2818  (1992).

\bibitem{VWN_LDA}
S.~H. Vosko, L. Wilk, and M. Nusair, Can.\ J.\ Phys. {\bf 58},  1200  (1980).

\bibitem{smith69}
J.~R. Smith, Phys.\ Rev.\ {\bf 181},  522  (1969).

\bibitem{HarrisArgument}
C. Yannouleas and U. Landman, Phys.\ Rev.\ B {\bf 48},  8376  (1993).

\bibitem{BerryMount}
M.~V. Berry and K.~E. Mount, Rep.\ Prog.\ Phys.\ {\bf 35},  315  (1972).

\bibitem{Tatievski94}
B. Tatievski, P. Stampfli, and K.~H. Bennemann, Z.\ Phys.\ D {\bf 31},  287
  (1994).

\bibitem{EXAFS1}
G. Apai, J.~F. Hamilton, J. Stohr, and A. Thompson, Phys.\ Rev.\ Lett.\ {\bf
  43},  165  (1979).

\bibitem{ToBePublished}
E. Koch, to be published.

\bibitem{StabilJelly}
J.~P. Perdew, H.~Q. Tran, and E.~D. Smith, Phys.\ Rev.\ B {\bf 42},  11627
  (1990).

\end{thebibliography}

\end{multicols}
\end{document}